# Hybridization-induced resonances with high quality factor in a plasmonic concentric ring-disk nanocavity

Zhaojian Zhang [1, *], Junbo Yang [2], Heng Xu [1], Siyu Xu [1], Yunxin Han [2], Xin He [2], Jingjing Zhang [1], Jie Huang [1] and Dingbo Chen [1]

[1] College of Liberal Arts and Sciences, National University of Defense Technology, Changsha 410073, China;
[2] Center of Material Science, National University of Defense Technology, Changsha 410073, China;
*Correspondence: 376824388@qq.com

**Abstract**
Plasmonic resonators have drawn more attention due to the ability to confine light into subwavelength scale. However, they always suffer from a low quality (Q) factor owing to the intrinsic loss of metal. Here, we numerically propose a plasmonic resonator with ultra-high Q factor based on plasmonic metal-insulator-metal (MIM) waveguide structures. The resonator consists of a disk cavity surrounded by a concentric ring cavity, possessing an ultra-small volume. Arising from the plasmon hybridization between plasmon modes in the disk and ring cavity, the induced bonding hybridized modes have ultra-narrow full wave at half maximum (FWHM) as well as ultra-high Q factors. The FWHM can be nearly 1 nm and Q factor can be more than 400. Furthermore, such device can act as a refractive index sensor with ultra-high figure of merit (FOM). This work provides a novel approach to design plasmonic high-Q-factor resonators, and has potential on-chip applications such as filters, sensors and nanolasers.

## 1. Introduction

In the past few decades, optical resonators have been playing an important role in photonic research. For fundamental research, they provide ideal platforms to study light–matter interaction [1], chaos [2] and nonlinear effects [3]. For applied research, they promote a series of significant applications including sensors [4], lasers [5], delay lines [6] and frequency comb [7]. In these applications, quality (Q) factor (which is proportional to the photon confinement time) and volume of the cavity figure prominently, since higher Q factors and smaller cavity volumes will enhance the performances of these devices [1]. However, the volumes of conventional dielectric and semiconductor cavities are limited to micron scales due to the diffraction limit [8].

Towards ultra-small volume devices, plasmonics is introduced to optical systems. Surface plasmons can confine visible and near-infrared light on subwavelength scales, thus bringing plasmonic resonators into nanoscales [9]. However, the Q factors of plasmonic resonators have been less than 100 both for visible and near-infrared wavelengths due the ohmic loss of the metal [10-12]. Plasmonic dielectric-metal hybrid resonators are introduced to increase the Q factor, but still possess footprints as large as several micrometers [13-14]. Recently, as a strong candidate to build high-compact optical circuits on chips, plasmonic metal-insulator-metal (MIM) waveguides are utilized to construct plasmonic resonators with ultra-small footprint [15-16]. For example, a square resonator is introduced based on MIM waveguides with Q factor 20, approximately [17]. Utilizing the Fano resonance, MIM-based resonator can reach a Q factor of 120 [18]. It has also been reported that a double-ring resonator can have a Q factor of nearly 300 [19]. Therefore, MIM waveguides have the potential to support high-Q-factor resonances in plasmonic nanocavities.

In this work, we introduce a resonator comprised of a concentric ring and disk cavity based on MIM

waveguides. Due to the plasmon hybridization, there is an interaction between plasmon modes both in ring and disk cavity. As a result, the induced bonding hybridized modes possess ultra-narrow full wave at half maximum (FWHM) and ultra-high Q factors in visible band. The influences of gap widths are also studied. After structural optimization, this resonator can obtain a FWHM nearly 1 nm and a Q factor more than 400. As a sensor, this device has considerable sensitivity and ultra-high figure of merit (FOM). Such resonator has an ultra-small volume and can find on-chip applications on sensing, filtering, lasing and nonlinear enhancement.

## 2. Results and discussion

The 2D schematic of proposed structure is shown in Fig. 1, which is constructed by an input waveguide and a side-coupled concentric ring-disk resonator. Here, the grey color represents silver (Ag), and white color represents air. The geometric parameters are provided in the caption of Fig. 1. The dispersive permittivity of Ag is adapted following the Drude model [20]:

$$\varepsilon_m = \varepsilon_\infty - \frac{\omega_p^2}{\omega(\omega+i\gamma)} \tag{1}$$

Where $\varepsilon_\infty$ is the permittivity of infinite frequency, $\omega_p$ is bulk plasma frequency, $\gamma$ is electron oscillation damping frequency, and $\omega$ is angular frequency of incident light. The parameters of Ag are as follows: $\varepsilon_\infty=3.7$, $\omega_p= 1.38\times10^{16}$ Hz, and $\gamma= 2.73\times10^{13}$ Hz.

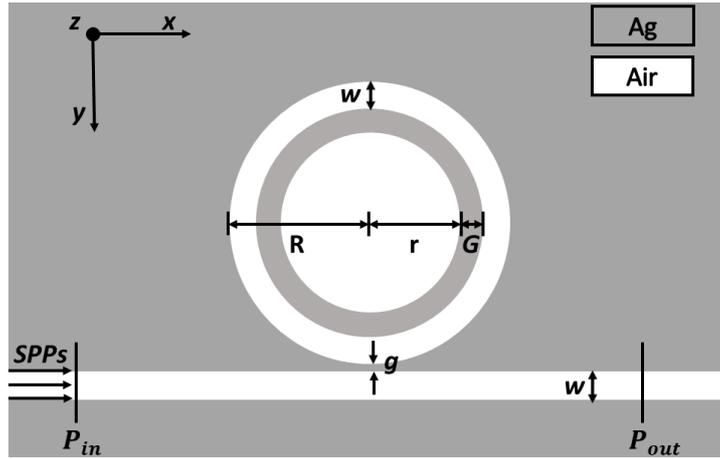

Fig. 1 The 2D schematic of proposed structure. The geometric parameters are as follows: $w$= 50 nm, $g$= 10 nm, $G$= 34 nm, $R$= 325 nm, and $r$= 241 nm.

There is only transverse-magnetic (TM) mode in the MIM waveguide [21]. Since the waveguide width $w$ is much smaller than input wavelengths, only fundamental TM mode can be supported. The dispersion of this mode is as follows [22]:

$$\frac{\varepsilon_i p}{\varepsilon_m k} = \frac{1-e^{kw}}{1+e^{kw}}$$

$$k = k_0\sqrt{(\frac{\beta_{spp}}{k_0})^2 - \varepsilon_i}, p = k_0\sqrt{(\frac{\beta_{spp}}{k_0})^2 - \varepsilon_m} \tag{2}$$

$$\beta_{spp} = n_{eff}k_0 = n_{eff}\frac{2\pi}{\lambda}$$

Here, $\lambda$ is incident wavelength in vacuum, $\varepsilon_i$ and $\varepsilon_m$ are dielectric and metal constant, respectively. $\beta_{spp}$ is propagation constant of surface plasmons, $n_{eff}$ is effective refractive index of MIM waveguide, and $k_0 = 2\pi/\lambda$ is wave number. 2D Finite-Difference Time-Domain (FDTD) method is adapted to do the simulation since it has the same result as 3D simulation when the height (in the z-direction) of this structure is set as 1 um [23]. The mesh size is 2.5 nm, the boundary condition is perfectly matched layers (PML). As depicted in Fig. 1, light source and power monitor is placed at $P_{in}$, while another power monitor is placed at $P_{out}$. The transmission is defined as $T = P_{out}/P_{in}$.

We start with the simple structure including an input waveguide side-coupled with a single disk resonator (the radius is 241 nm and the gap width between waveguide and disk is 34 nm), or a single ring resonator (the radius is 325 nm and the gap width is 10 nm). The resonance conditions are as follows [15, 24]:

$$\text{Disk}: k_d \frac{H_m^{(1)'}(k_e r)}{H_m^{(1)}(k_e r)} = k_e \frac{J_m'(k_d r)}{J_m(k_d r)}, m = 1, 2...$$

$$\text{Ring}: n\lambda = L_{eff} \, \text{Re}(n_{eff}), n = 1, 2...$$

(3)

For the disk, $r$ is the disk radius, $k_d$ and $k_e$ are the wave vectors in the disk and metal, $H_m^{(1)}$ and $H_m^{(1)'}$ are the first kind Hankle function with order $m$ and its derivative, $J_m$ and $J_m'$ are the first kind Bessel function with order $m$ and its derivative, respectively. For the ring, $L_{eff}$ is effective perimeter of the ring, which is the average of the inner and outer perimeters, $\lambda$ is the resonant wavelength. Both $m$ and $n$ are integer mode numbers. The corresponding transmission spectra are shown in Fig. 2(a). For the disk resonator, there are two resonances at 446.3 nm and 586.5 nm, FWHM of each is 1.7 and 1.9 nm, and Q factor, defined as $Q = \lambda/\text{FWHM}$, is 262.5 and 308.7, respectively. According to the $H_z$ field distributions in insets of Fig. 2(a), $m$= 3 and 2 for disk modes at 446.3 nm and 586.5 nm, respectively. For the two ring modes at 466 nm and 546.7 nm, $n$= 6 and 5, respectively.

Fig. 2(b) shows the transmission spectrum of the concentric ring-disk resonator given in Fig. 1, the mode numbers and field distributions are provided in insets of Fig. 2(b). Basically, this spectrum is a combination of disk and ring spectrum. However, there are two interesting phenomena. Firstly, there are two types of plasmon modes in such hybrid resonator: antisymmetrically coupled (antibonding) modes (mode III at 472.6 nm and mode IV at 554.1 nm), and symmetrically coupled (bonding) modes (mode I at 450.7 nm and mode II at 590.7 nm). Interestingly, the energy is mainly confined in the outer ring for antibonding modes and inner disk for bonding modes. Such plasmon mode distributions can be described by the plasmon hybridization [25-27], the corresponding energy diagram is shown in Fig. 3. According to this model, the plasmon hybridization between two plasmon sub-modes will gives rise to two hybridized plasmon modes: the antibonding mode and bonding mode. Therefore, such hybridization process will happen via the coupling between the disk and ring nanocavity, leading to antibonding hybridized modes with lower energy (mode III and IV) and bonding hybridized modes with higher energy (mode I and II) in this system.

Secondly, bonding modes have ultra-narrow FWHM and ultra-high Q factors. We believe this is due to the special mode distributions of bonding modes. The energy of such modes is mainly trapped in the inner disk nanocavity, so the outer ring nanocavity can suppress the dissipation rate of photons confined to the disk nanocavity. In this case, FWHM and Q factor for mode I are 1.1 nm and 417.3, and for mode II are 1.4 nm and 418.9, respectively. Such values are better than previous similar works [15-19].

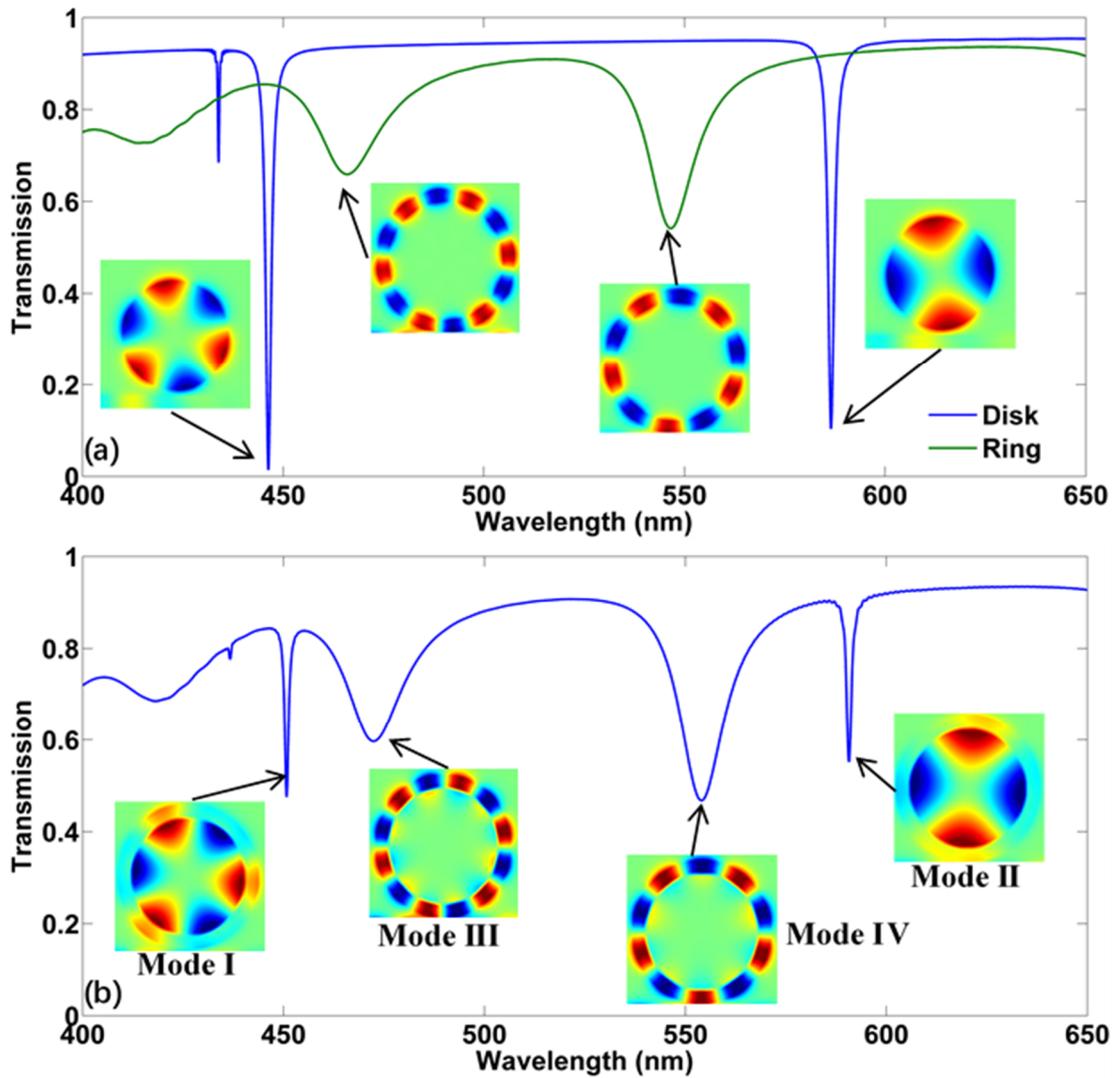

Fig. 2 (a) The transmission spectra of single disk and ring resonator. The insets show $H_z$ distributions of each mode. (b) The transmission spectrum of the concentric ring-disk resonator. The insets show $H_z$ distributions and numbers of each mode.

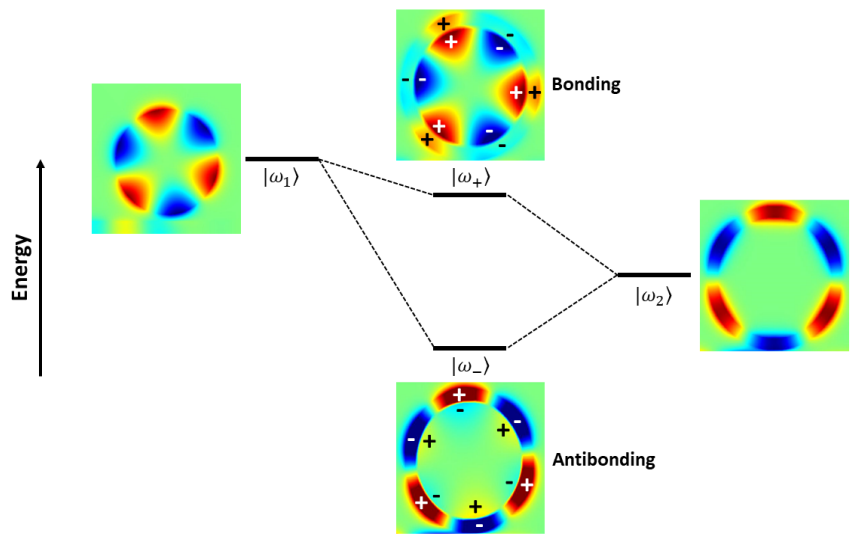

Fig. 3 The energy diagram of plasmon hybridization in this system.

Next, the influence of gap width will be studied, including the *g*, which is the gap width between the input waveguide and ring-disk resonator, and *G*, which is the gap width between ring and disk nanocavity. As shown in Fig. 4. As *g* increases (keep *G* unchanged at 34 nm), the transmission at each hybridized mode will increase, indicating that there will be less power coupled into the resonator. The same is true when *G* increases (keep *g* unchanged at 10 nm and reduce *r*). This is because wider metal gap will lead to more ohmic loss. Furthermore, increasing *g* will cause a resonant wavelength blueshift of mode III and IV, but will not for mode I and II as presented in Fig. 4(a). This is because, altering coupling distance will bring about the phase change in the coupling process, leading to the coupling-induced resonant frequency shift (CIFS) [28]. Resonant wavelengths of antibonding/bonding modes are mainly decided by ring/disk nanocavity, so *g* will directly influence resonant wavelengths of antibonding modes and will not for bonding modes. From the same fact, *G* will affect the resonant wavelengths of both hybridized modes as shown in Fig. 4 (b).

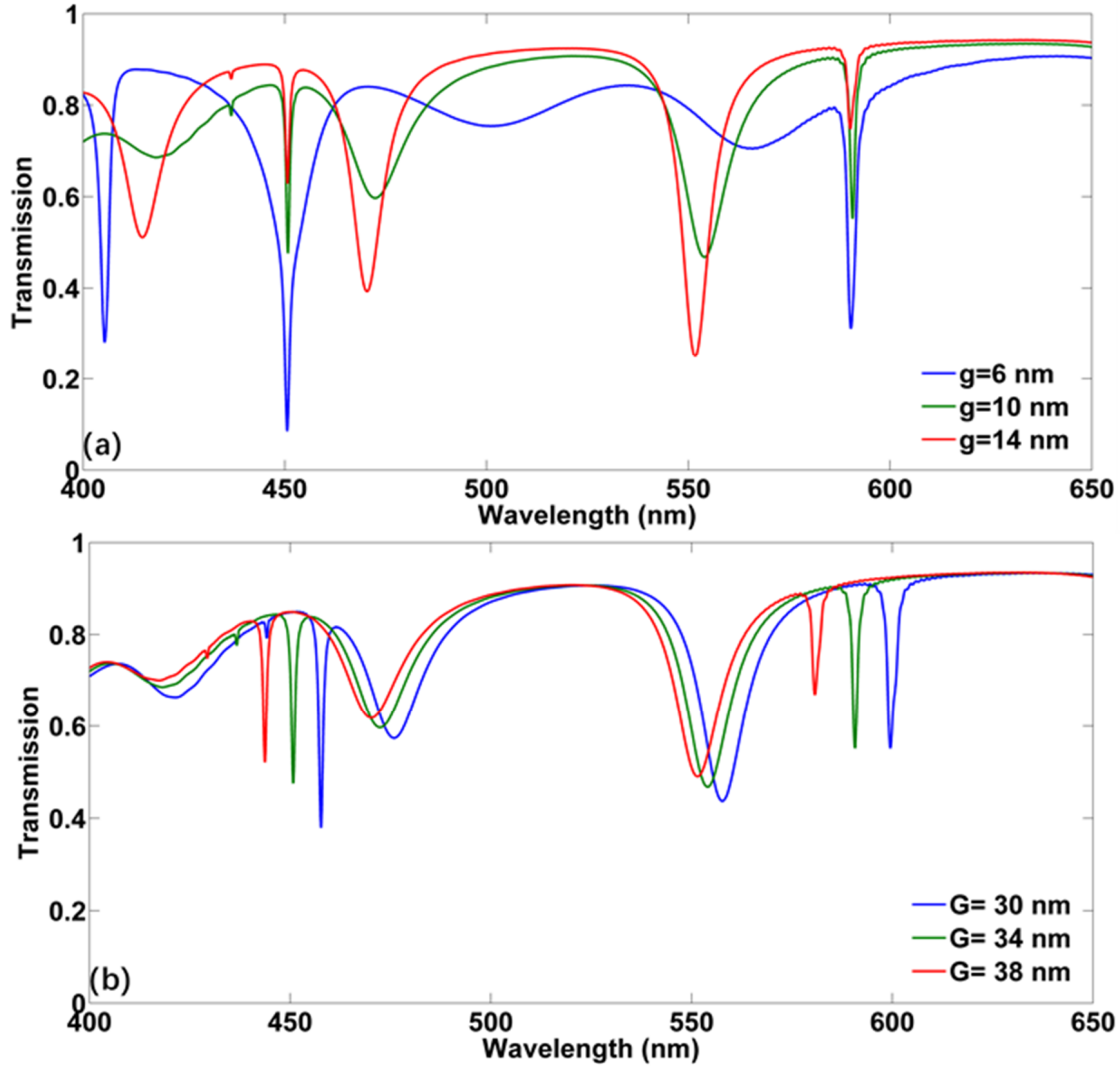

Fig. 4 (a) The transmission spectra under different *g*. (b) The transmission spectra under different *G*.

Due to the superior energy storage capacity, bonding modes will be focused next, and two important factors of which will be investigated under different gap widths: Q factor, which represents energy storage time, and contraction ratio, defined as $T_c = T_{max}/T_{min}$ ($T_{max}$ and $T_{min}$ are the transmission at the highest and lowest point of resonance dips, respectively), indicating the amount of stored energy inside. The relationship diagrams are given in Fig. 5. From both diagrams we can see that increasing gap widths will lead to a decline of contraction ratio, which arising from the intrinsic loss of metal gap. However, Q factor will rise on the main trend, because wider gap will promote photon confinement time. Therefore, there is a trade-off between Q factor and contraction ratio.

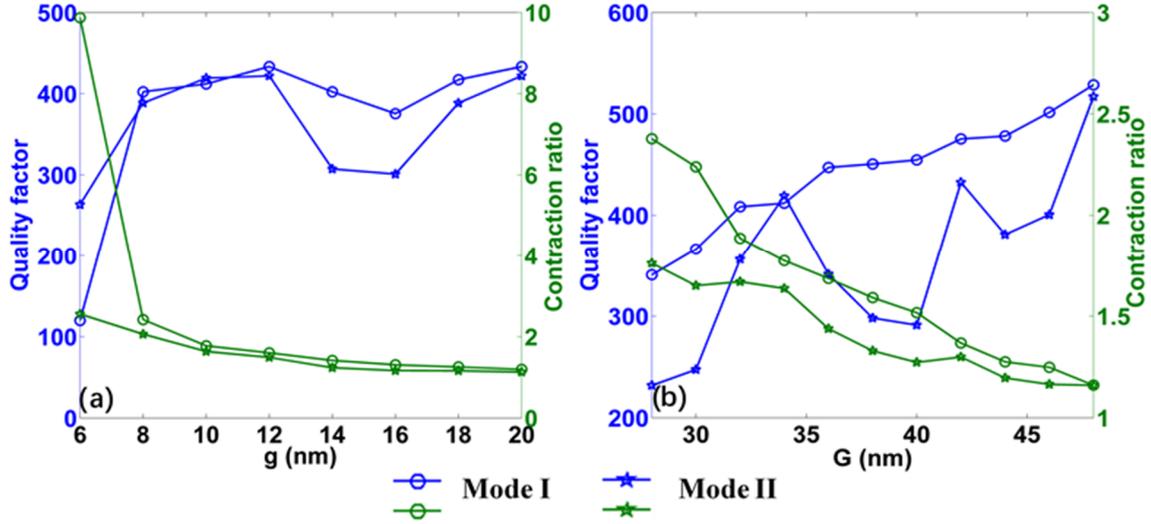

Fig. 5 (a) The relationship diagram between *g* and Q factor/contraction ratio of mode I and II. (b) The relationship diagram between *G* and Q factor/contraction ratio of mode I and II.

## 3. Sensing performance

In consideration of the balance between Q factor and contraction ratio, we adopt the structure with *g*= 10 nm and *G*= 34 nm to investigate the corresponding sensing performance. A sensor can be assessed by two factors, sensitivity (S) and FOM [19]:

$$S=\frac{\Delta \lambda}{\Delta n}$$
$$FOM = \frac{S}{FWHM} \quad (4)$$

Here, S represents the wavelength shift induced by unit change of surrounding refractive index, FOM indicates the optical resolution of the sensor. The resonant wavelengths and FWHM of mode I and II under different surrounding refractive index are shown in Fig. 6 (a) and (b), respectively, from which we can see that the resonant wavelengths of both modes have linear relationships with surrounding refractive index. Therefore, S of mode I is 429 nm/RIU, and of mode II is 579 nm/RIU. The average FWHM of mode I and II under surrounding refractive index from 1 to 1.1 is 1.1 nm and 1.5 nm respectively, so FOM of mode I is 376.2/RIU, and of mode II is 378/RIU. Compared with other MIM-based sensors, this sensor has moderate sensitivity since the working wavelength is in the visible band, which is due to the ultra-small footprint of this sensor. However, this sensor has much higher FOM than other similar previous works [19, 29-31], therefore possessing ultra-high optical resolution.

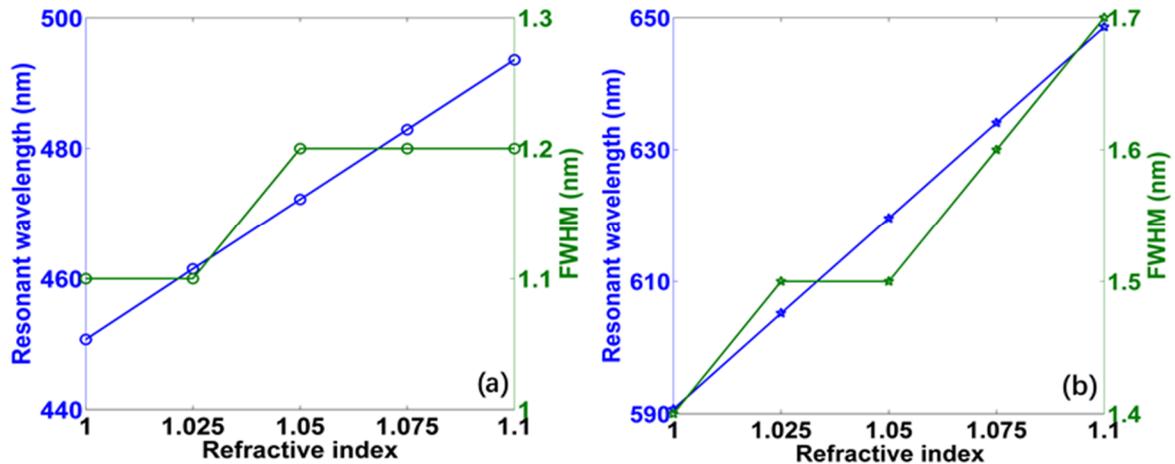

Fig. 6 (a) The resonant wavelengths and FWHM of mode I under surrounding refractive index from 1 to 1.1. (b) The resonant wavelengths and FWHM of mode II under surrounding refractive index from 1 to 1.1.

**4. Conclusion**

In summary, we propose a plasmonic hybrid ring-disk nano resonator with high Q factors and small footprint based on MIM waveguides. Due to plasmon hybridization, bonding and antibonding hybridized plasmon modes are generated, and ultra-narrow FWHM and ultra-high Q factors can be supported by bonding modes. The performance as a sensor with ultra-high FOM is also investigated. This work provides a way to improve the Q factor by using hybridized modes, and can be applied on light-matter interaction, sensing and nanolasers on chipscales.

**Acknowledgments**

This work is supported by the National Natural Science Foundation of China (61671455, 61805278), the Foundation of NUDT (ZK17-03-01), the Program for New Century Excellent Talents in University (NCET-12-0142), and the China Postdoctoral Science Foundation (2018M633704).